\documentclass[aps,prl,twocolumn,showpacs,superscriptaddress,floatfix,nofootinbib]{revtex4}

\usepackage{latexsym}
\usepackage{graphicx}
\usepackage{epsfig}
\usepackage{amssymb}
\usepackage[english]{babel}

\begin{document}

\title{Two Slightly-Entangled NP-Complete Problems}

\author{Rom\'an Or\'us}
\affiliation{Dept. d'Estructura i Constituents de la Mat\`eria,
   Univ. Barcelona, 08028, Barcelona, Spain}

\begin{abstract}
We perform a mathematical analysis of the classical computational
complexity of two genuine quantum-mechanical problems, which are 
inspired in the calculation of the expected magnetizations and the
entanglement between subsystems for a quantum 
spin system. These problems, which we respectively call SES and
SESSP, are specified in terms of pure slightly-entangled
quantum states of $n$ qubits, and rigorous mathematical
proofs that they belong to the NP-Complete complexity class are presented. Both 
SES and SESSP are, therefore, computationally equivalent to the relevant $3$-SAT problem, 
for which an efficient algorithm is yet to be discovered. 

\end{abstract}

\pacs{03.65.Ud, 03.67.Lx}

\maketitle

The classification of different problems in terms of the
computational resources required to solve them, and the existing interrelations
between them, is the scope of the field of computational complexity
\cite{papadimitrou, gareyjohnson}. This can often provide useful information 
for algorithm designers in order to develop tools to attack 
particular problems. If a problem is known to be ``easy'', then it should 
be easy to design an algorithm to solve it, whereas if a problem is known to be
``hard'', the search for a possible efficient algorithm should require of a great
effort, and the existence of such an algorithm is not always guaranteed. 

In 1971, S. Cook \cite{cook} proved a significant mathematical result, namely,
that all the problems in the complexity class NP \cite{NPnote}
can be efficiently mapped to the problem of determining whether a
given set of clauses, each one of them involving $k$ literals of boolean variables,
was satisfiable or not, for $k=3$. This problem, known as $3$-SAT,
is an NP-Hard problem, since all NP problems can be efficiently reduced to it,
but furthermore it is also a member of the class NP, and therefore
it becomes a genuine representative of the whole class. Solving
$3$-SAT with some resources immediately solves
the whole NP complexity class without basically changing these 
resources. The class of problems sharing this property, namely, that they are
both NP-Hard and NP, are the so-called NP-Complete problems, and the existence
or not of an efficient algorithm for them lies at the heart of the celebrated 
P=NP conjecture \cite{PNP}, being therefore of great theoretical importance. On top, since
the pioneering work of Cook, there has been an explosion in the identification 
of new NP-Complete problems belonging to very different fields. Nowadays,
these problems are known to appear in situations such as graph theory, network design,
storage and compression of information, multiprocessor scheduling, solvability
of equations, automata theory and chip verification, among many
others. The relevance of NP-Complete problems is consequently not only
mathematical, but also practical for current technology. 

Because of their importance, the search of algorithms for these
problems has become a topic of intense research. So far, no -classical
or quantum- algorithm has been proven to be efficient when dealing with this
class of problems, and the existence or not of such an algorithm remains
yet as a relevant open question. Nevertheless, recent results in quantum
computation \cite{NielsenandChuang} seem to indicate that quantum algorithms
might be more powerful than their classical counterparts in solving NP-Complete
problems \cite{Farhi1, Farhi2, Farhi3, Orus1, Orus2}. It is a
well-known fact that
quantum mechanics offers powerful computational capabilities because of
the existence of entanglement \cite{Jozsa, Vidal1, Vidal2} and, if
quantum computation is to bring more efficient algorithms for solving NP-Complete
problems, a first natural step in this direction is to state these problems in terms of the
genuine quantum-mechanical language. The discovery of new NP-Complete 
problems in quantum mechanics is, therefore, a major task towards the development of new 
quantum algorithms to solve them. Additionally, while many 
problems are known to belong to the NP-Complete complexity class, still 
very few problems have been discovered to belong to its quantum analogue, the so-called 
QMA-Complete complexity class \cite{Kitaev1, Aharonov, Kempe1, Kempe2,
  Qcircuit}. The analysis of the classical complexity of quantum problems
is also a natural link between the classical and quantum complexity theories, which
might be helpful in order to bring new QMA-Complete problems. 

In this letter we present two genuine quantum-mechanical problems
which we strictly prove to be NP-Complete, and which are seen to be related to
the calculation of the expected magnetizations and the entanglement between
subsystems for a quantum spin system. For these two problems, entanglement is
seen to play the role of a tuning parameter between complexity classes, with
increasing level of the needed computational resources as quantum correlations 
increase. Let us initially define what we understand by a $n$-qubit slightly-entangled
pure quantum state, and by a weight function for mixed
quantum states. The first of these definitions reads as follows: 

\vspace{5pt}

{\bf Definition 1:} a \emph{$n$-qubit slightly-entangled pure quantum state}
$|\psi\rangle$  is a state such that the maximum of all the ranks of the reduced density
matrices obtained from all the possible bipartitions of the system,  
$\chi$, grows at most polynomially with
the number of qubits, that is, $\chi = O({\rm poly}(n))$. 

\vspace{5pt}

This definition was initially introduced by Vidal in \cite{Vidal1} in order to
propose an efficient classical simulation protocol for quantum computations
which only handle these -essentially little entangled- states, since $\chi$
is a measure of the maximum bipartite entanglement present in the system. The important 
point for us is that, according to \cite{Vidal1},
slightly-entangled states can efficiently be stored in $O({\rm poly}(n))$ space,
and the evaluation of local observables over them can be performed in
$O({\rm poly}(n))$ time and space. The second definition that we need is the
following:  

\vspace{5pt}

{\bf Definition 2:} a \emph{weight function $W$ for mixed quantum states}
is a mapping from the set of density matrices to the real numbers such
that {\it (i)} given a density matrix $\rho$ with rank $\chi_{\rho}$, $W(\rho)$ can be computed in 
$O({\rm poly}(\chi_{\rho}))$ time and space, {\it (ii)} $W(\rho_1 \otimes \rho_2 ) = W(\rho_1) +  
W(\rho_2)$, and {\it (iii)} $0 \le W(|a\rangle \langle a|) \le 1$ for a single-qubit pure quantum
  state $|a\rangle$.

\vspace{5pt}

The concept of weight function is actually rather general, and there are many quantities
satisfying the above requirements. We wish to remark that, by definition, a
weight function $W$ is computable in polynomial time in the rank of the
density matrix $\rho$ to which it is applied. This will turn to be a key ingredient
in our proofs of NP-completeness that we shall present below.   
A practical example of a weight function for $m$-qubit density matrices is the
expected value of a sum of
single-qubit observables which are positive semidefinite. For example, the
expected value of the operator $\hat{X}^{\alpha} = \sum_{i=1}^m
\frac{1}{2}(1-\hat{\sigma}^{\alpha}_{i}) = \frac{m}{2}-\hat{S}^{\alpha}$,
$\hat{\sigma}_i^{\alpha}$ and $\hat{S}^{\alpha}$ respectively being the Pauli 
matrix for qubit $i$ and the total spin operator, both in direction
$\alpha$. In fact, this weight function is basically the magnetization in the direction
$\alpha$ of the sample of $m$ spins $1/2$. Another totally different example
of weight function for density matrices $\rho$ is the Von Neumann entropy 
$S(\rho) = - {\rm tr}(\rho{\rm log}_2 (\rho))$. Since the Von Neumann entropy
is a measure of the bipartite entanglement present in pure quantum states
\cite{NielsenandChuang}, we see that the generic concept of weight function
naturally connects, as a particular case, with the ideas of quantification 
of the quantum correlations present in a given quantum system.

At this point we define the decision problems we wish to focus on, which
we shall respectively call SES problem (from Slightly-Entangled State) and
SESSP problem (from Slightly-Entangled State Splitting): 

\vspace{5pt}

{\bf SES problem:}

{\it Instance:} an $n$-qubit slightly-entangled pure quantum state $|\psi\rangle$, $2$ positive real
numbers $B$, $\epsilon$, such that $B > \epsilon$, and a weight function $W$ 
for mixed states. All real numbers are expressed with 
$O({\rm poly}(n))$ precision.

{\it Question:} is there some subset -characterized by a
density matrix $\rho$- of the $n$-qubit system such that 
$B-\epsilon \le W(\rho) \le B+\epsilon$?

\vspace{5pt}

{\bf SESSP problem:}

{\it Instance:} an $n$-qubit slightly-entangled pure quantum state
$|\psi\rangle$, a positive real
number $\epsilon$, and a weight function $W$ for mixed states. 
All real numbers are expressed with 
$O({\rm poly}(n))$ precision.

{\it Question:} is there some splitting -characterized by 
density matrices $\rho_{A_1}$ and $\rho_{A_2}$, $A_1$ and $A_2$ being the subsystems of
the bipartition- of the $n$-qubit system such that $W(\rho_{A_1}) = W(\rho_{A_2}) \pm \epsilon$?

\vspace{5pt}

We wish to stress that the instances of the SES and SESSP problems are
entirely defined in terms of quantum-mechanical quantities. Furthermore, because
of definitions $1$ and $2$, all the possible instances of SES and SESSP can be described
with the use of $O({\rm poly}(n))$ space, $n$ being the number of qubits in
the system. This is a necessary requirement if we wish the problems to have
some possibility of being solvable by an efficient algorithm, namely, that their
description can not consume exponential resources. In our case, a key
point for this is the fact that the $n$-qubit pure quantum state we deal with is only 
slightly-entangled. Our aim now is to analyze 
the classical complexity of the SES and SESSP problems, and 
for that we prove two theorems, which respectively state that both SES and SESSP are
NP-Complete. We develop in detail the proof of NP-Completeness of SES, and
the proof of NP-Completeness of SESSP will follow exactly the same ideas. For
the first of these problems, we have the following result: 

\vspace{5pt}

{\bf Theorem 1:} SES $\in$ NP-Complete. 

{\it Proof:} in order to prove that SES $\in$ NP-Complete, we have to prove
that {\it (i)} SES $\in$ NP, and that {\it (ii)} SES $\in$ NP-Hard. This is
the purpose of the following two propositions:

{\it (i) Proposition 1:} SES $\in$ NP. 

{\it Proof}: since the $n$-qubit pure state of the instance is only
slightly-entangled, it can be described by means of $O({\rm poly}(n))$
space. In this case, the certificate -or witness- is simply the set of qubits which verify
that $B-\epsilon \le W(\rho) \le B+\epsilon$. Given this set of qubits, $\rho$
can be computed in $O({\rm poly}(n))$ time and described by $O({\rm poly}(n))$
coefficients, since the pure quantum state of the whole system is
slightly-entangled. By definition, $W(\rho)$ can then be computed in
polynomial time. Therefore, given an instance to the problem, a
possible certificate can be verified in polynomial time only by classical
means. This is precisely the definition of the complexity class NP, so we
conclude that SES $\in$ NP. $\Box$ 

{\it (ii) Proposition 2:} SES $\in$ NP-Hard.

{\it Proof}: we prove this point by reduction of all the instances of a known
NP-Complete problem to particular instances of the SES problem. The problem we
make use of, and which is known to be NP-Complete, is called SUBSET SUM
\cite{gareyjohnson} and reads as follows: 

\vspace{5pt}

SUBSET SUM problem:

{\it Instance:} finite set $A$, size $s(a) \in {\mathbb Z}^+ \ \forall a \in
A$, $B \in {\mathbb Z}^+$. 

{\it Question:} is there some subset $A' \in A$ such that $\sum_{a \in A'}
s(a) = B$?

\vspace{5pt}

We are going to reduce every possible instance of SUBSET SUM to a particular
instance of SES. This will prove that SES is, at least, as hard as SUBSET SUM,
and therefore SES will be proven to be NP-Hard. The first step is to perform 
an extension of SUBSET SUM to the real domain as follows:

\vspace{5pt}

REAL SUBSET SUM problem:

{\it Instance:} finite set $A$, size $s(a) \in {\mathbb R}^+\cup\{0\} \ \forall a \in
A$, $B \in {\mathbb R}^+\cup\{0\}$, $\epsilon \in {\mathbb R}^+\cup\{0\}$, $B
> \epsilon$. All real numbers are expressed with $O({\rm poly}(|A|))$
precision, $|A|$ being the size of the set $A$. 

{\it Question:} is there some subset $A' \in A$ such that $B-\epsilon \le \sum_{a \in A'}
s(a) \le B+\epsilon$?

\vspace{5pt}

It is straightforward to see that all the instances of SUBSET SUM can be
reduced to particular instances of REAL SUBSET SUM. Therefore, REAL SUBSET SUM
is an NP-Hard problem. Now, we redefine each one of the instances of this
problem by means of a normalization factor. Calling $C = \sum_{a
  \in A} s(a)$, we perform the following changes for $C \ne 0$: $\tilde{s}(a) \equiv
s(a)/C$, $\tilde{B} \equiv B/C$, and $\tilde{\epsilon} \equiv \epsilon/C$. The case
$C=0$ corresponds to the particular instance in which $s(a)=0 \ \forall a \in
A$, which we leave unaltered. Then, the following problem is also NP-Hard: 

\vspace{5pt}

REAL SUBSET SUM 2 problem:

{\it Instance:} finite set $A$, size $\tilde{s}(a) \in {\mathbb
  R}^+\cup\{0\}$, $0 \le \tilde{s}(a) \le 1 \ \forall a \in A$, $\tilde{B} \in 
{\mathbb R}^+\cup\{0\}$, $\tilde{\epsilon} \in {\mathbb R}^+\cup\{0\}$, $\tilde{B}
> \tilde{\epsilon}$, and $\tilde{B} \le |A|$. All real numbers are expressed 
with $O({\rm poly}(|A|))$ precision, $|A|$ being the size of the set $A$.  

{\it Question:} is there some subset $A' \in A$ such that
 $\tilde{B}-\tilde{\epsilon}  \le \sum_{a \in A'} \tilde{s}(a) \le
 \tilde{B}+\tilde{\epsilon}$?

\vspace{5pt}

This problem is still NP-Hard, since all the instances of REAL SUBSET SUM
correspond to particular instances of REAL SUBSET SUM 2 (note that in this
last problem $\sum_{a \in A}\tilde{s}(a)$ can be any number). 
The mapping between the instances of REAL SUBSET
SUM and of REAL SUBSET SUM 2 is not one-to-one, since all the instances 
from REAL SUBSET SUM differing by an overall multiplicative constant are
mapped to the same instance of REAL SUBSET SUM 2. All these instances turn out
to be equivalent from the computational point of view. With the normalization we have
removed the redundancy in the instances of REAL SUBSET SUM, mapping all the
equivalent ones to a representative of its equivalence class, the
equivalence relation being defined by the overall multiplication by a constant. 

Now, we associate to each instance of REAL SUBSET SUM 2 the
following instance of SES: a $n$-qubit product pure quantum state
$|\psi\rangle = |a_1\rangle
|a_2\rangle \cdots |a_n\rangle$, a weight function satisfying
$W(|a_i\rangle \langle a_i|) = \tilde{s}(a_i) \ \forall i = 1, \ldots, n$, and
the number of qubits being $n = |A|$. Solving a given instance of REAL SUBSET 
SUM 2 is then equivalent to solving a particular instance of SES for a 
particular weight function and for a separable pure quantum state. 
Therefore, SES is NP-Hard. $\Box$

Since we have both proved that SES $\in$ NP and that SES $\in$ NP-Hard, we
conclude that SES $\in$ NP-Complete. $\Box$

For the second of the problems we can prove an analogue result to the one
presented above. It reads as follows:

\vspace{5pt}

{\bf Theorem 2:} SESSP $\in$ NP-Complete. 

{\it Proof:} the proof of this theorem follows exactly the same lines of the
proof of the NP-Completeness of SES, so we will not develop again all the
particular details, since they have already been presented in
the previous theorem. Once more, in order to prove that SESSP $\in$
NP-Complete, we need to prove that {\it (i)} SES $\in$ NP, and that 
{\it (ii)} SES $\in$ NP-Hard. The proof that SESSP $\in$ NP is exactly
equivalent to the proof in proposition 1 that SES $\in$ NP, with the only
exception that the certificate in
this case is the list of qubits belonging to each one of the settings of the
bipartition. From this information everything can be verified with the use of polynomial
resources, exactly as in proposition 1, therefore SESSP $\in$ NP. For proving
that SESSP $\in$ NP-Hard, we do again a reduction of all the instances of a
known NP-Complete problem to particular instances of the SESSP problem. In
this case, the NP-Complete problem from which we perform the reduction is 
called PARTITION \cite{gareyjohnson}, and is defined as follows:

\vspace{5pt}

PARTITION problem:

{\it Instance:} finite set $A$, size $s(a) \in {\mathbb Z}^+ \ \forall a \in
A$. 

{\it Question:} is there some subset $A' \in A$ such that $\sum_{a \in A'}
s(a) = \sum_{a \in (A-A')}s(a)$?

\vspace{5pt}

Exactly in the same way as we proceeded in the proof of proposition 2, the
reduction from PARTITION to SESSP is made initially by a generalization of
PARTITION to the real domain (REAL PARTITION), then by a normalization of all the
equivalent instances (REAL PARTITION 2), and finally by an association of the
instances of this last problem with instances of SESSP over separable states, 
in the same way as we already performed in the last step of the proof of
proposition 2. This proves the NP-Hardness of SESSP, and since SESSP is also
proved to be NP, we conclude that SESSP $\in$ NP-Complete. $\Box$

\vspace{5pt}

We would like to point out the role entanglement plays in the previous
proofs of NP-Completeness. We have seen in the proofs that, in order for SES
and SESSP to be NP-Complete problems, we can restrict ourselves to instances dealing with
separable pure quantum states of $n$ qubits, for which $\chi = 1$. Many of the
instances dealing with pure quantum states with $\chi = O(2^n)$ can not
probably be defined as instances of an NP problem, since they may require of
exponential resources for their description. But if $\chi = O({\rm poly}(n))$,
then the polynomial description is guaranteed, and both SES and SESSP are still NP-Complete,
despite the significant existence of quantum correlations. This fact might be seen
as the analogous of the equivalence between classical
computation and slightly-entangled quantum computation, made explicit by Vidal in
\cite{Vidal1}, but now at the level of the complexity class
NP-Complete. Classical problems directly map to separable instances, and the presence
of some entanglement in the system still does not change the classical
complexity class of the problem.   

We also note that there can be particular sets of instances of SES
and SESSP with a particular weight function $W$ which are already NP-Complete on their own. As a
practical example, we can restrict ourselves to the instances in which the
weight function $W$ is basically the expected value of the magnetization of a
sample of spins $1/2$, in the way already described after definition 2. It is easy to prove, 
following the same procedures as before, that even with this restriction 
both the SES and SESSP problems are still NP-Complete problems. For instance, 
knowing whether there is a set of magnetic domains in a 
slightly-entangled quantum spin system, 
such that the magnetic domains on the whole have a particular total value 
of the magnetization, is an NP-Complete problem. Another remarkable set of
instances which are easily seen to be NP-Complete for SES but not for SESSP is defined 
when we choose the weight function to be the Von Neumann
entropy. The instances defined in this way are indeed trivial instances of
SESSP, since the Von Neumann entropy is always identical for the two
subsystems in consideration regardless of the chosen partition, but they are
non-trivial instances of SES, since this value of the entropy can be 
partition-dependent. In fact, we can see that this set of instances is 
already NP-Complete for SES, by mapping all the possible instances of REAL SUBSET SUM
2 of size $|A|$ to an instance of SES defined by an appropiate $2n$-qubit pure quantum state $|\psi
\rangle = |\phi_{1,2}\rangle|\phi_{3,4}\rangle \cdots |\phi_{2n-1,2n}\rangle$,
$|\phi_{i,i+1}\rangle$ being entangled states of the qubits $i$ and $i+1$ such
that $S(\rho_i) = S(\rho_{i+1}) = \tilde{s}(a_i)$, $i = 1,3,5 \ldots
,2n-1$, and $n=|A|$. Note that at this point we have doubled the number of
qubits in the reduction, so more precisely we find that solving SES, defined for $2n$
qubits and in terms of the entanglement entropy, implies solving NP-Complete
problems for $n$ bits. Allowing a polynomial relation in the inputs, we conclude
that this set of instanes of SES is also NP-Complete, as claimed.  
Therefore, knowing how much entanglement is present in
a slightly entangled quantum spin system, is as hard as the $3$-SAT problem. 
We wish to note as well that slightly-entangled 
states naturally appear in one-dimensional
quantum spin systems, also called quantum spin-chains, where $\chi$ grows 
polynomially at the critical points, while being
constant away from criticality \cite{spinchain1, spinchain2, korepin,
  calabressecardy}. These models constitute interesting instances of
NP-Complete problems, which can be entirely stated in terms of suitable
quantum-mechanical quantities. 

A further study of the results 
of this paper would be of interest, 
trying to relate the presented NP-Complete problems to other intrinsic 
quantum-mechanical questions. In fact, a completely classical complexity 
analysis of many interesting
quantum problems is still to be developed, despite some results have already
been achieved, for example, in the separability problem \cite{gurvits}. A 
classical complexity study of quantum problems might be helpful 
in order to determine also their classification in terms of
quantum complexity theory, and should provide as well clear inspiration for the
design of new quantum algorithms. 

\vspace{5pt}

{\bf Acknowledgments:} the author is grateful to discussions with
J.I. Latorre, Y. Omar and N. Cerf about the content of this paper, and to financial
support from projects MCYT FPA2001-3598, GC2001SGR-00065 and IST-199-11053. 

{}


\begin{thebibliography}{}
\bibitem{NPnote}
NP is defined as the class of decision problems 
whose instances can be solved in polynomial time by a nondeterministic classical
computer. These problems have also the property that their instances 
can always be verified in polynomial time once a candidate solution is proposed.
\end{thebibliography}
\end{document}